\documentclass[a4paper,11pt]{article}
\pdfoutput=1 

\usepackage{jheppub} 

\usepackage[T1]{fontenc} 

\usepackage{amsmath,amssymb,epsfig,ulem,color,slashed,wasysym}
\allowdisplaybreaks  

\def \beq{\begin{equation}}
\def \eeq{\end{equation}}
\def \bea{\begin{eqnarray}}
\def \eea{\end{eqnarray}}

\def\bm#1{\mbox{\boldmath$#1$\unboldmath}} 

\title{\boldmath 
An explicit $Z^\prime$-boson explanation \\ of the $B \to K^\ast \mu^+ \mu^-$ anomaly}

\author[1]{Rhorry Gauld,}
\author[2]{Florian Goertz}
\author[3]{and Ulrich Haisch}

\affiliation[1]{Physics Department,
    University of Oxford, \\ OX1 3PN Oxford, United Kingdom}

\affiliation[2]{Institute for Theoretical Physics, ETH Zurich, 8093 Zurich, Switzerland}

\affiliation[3]{Rudolf Peierls Centre for Theoretical Physics,
    University of Oxford, \\ OX1 3PN Oxford, United Kingdom}

\emailAdd{r.gauld1@physics.ox.ac.uk}

\emailAdd{fgoertz@itp.phys.ethz.ch}

\emailAdd{u.haisch1@physics.ox.ac.uk}

\abstract{A global fit to the recent $B \to K^\ast \mu^+ \mu^-$ data shows indications for a large new-physics contribution to the Wilson coefficient of the semi-leptonic vector operator. In this article we consider a simple $Z^\prime$-boson model of $3$-$3$-$1$ type that can accommodate such an effect without violating any other constraint from quark-flavour physics. Implications for yet unobserved decay modes such as $B \to X_s \nu \bar \nu$ and longstanding puzzles like $B \to \pi K$ are also discussed. The $Z^\prime$-boson masses required to address the observed anomaly lie in the range of $7 \, {\rm TeV}$. Such heavy $Z^\prime$ bosons evade the existing bounds from precision data and direct searches, and will remain difficult to discover even at a high-luminosity  LHC. The potential of an ILC as well as the next generation of low-energy parity-violation experiments  in constraining the $Z^\prime$-boson parameter space is also examined.}

\begin{document} 

\maketitle

\flushbottom

\section{Introduction}

The success of the LHCb experiment has so far been a nightmare for all flavour physicists that were hoping to see signs of new physics popping up in $B_s$--$\bar B_s$ mixing and the rare $B_s \to \mu^+ \mu^-$ decay. This situation might have changed with the latest  measurements~\cite{Serra,Aaij:2013qta} of the angular correlations of the decay products  in $B \to K^\ast \mu^+ \mu^-$ that display several deviations from the standard model~(SM) predictions.  The largest discrepancy of $3.7 \sigma$ arises in the variable $P_5^\prime$~\cite{DescotesGenon:2012zf} (the analogue of $S_5$ in~\cite{Altmannshofer:2008dz}), which has specifically been designed to combine theoretical and experimental benefits, while retaining a high sensitivity to new-physics effects. Further LHCb analyses combined with a critical assessment  of theoretical uncertainties  (see in particular~\cite{Jager:2012uw} as well as very recently~\cite{Hambrock:2013zya}) will be necessary to clarify whether the observed deviations call for the presence of new physics or are simply  flukes.  

A first attempt to shed light on the $B \to K^\ast \mu^+ \mu^-$ anomaly has been made   in~\cite{Descotes-Genon:2013wba} (with subsequent new-physics analyses presented in~\cite{Altmannshofer:2013foa, Gauld:2013qba,Buras:2013qja}). By performing a global analysis of $B \to K^\ast \mu^+ \mu^-$, $B \to X_s \ell^+ \ell^-$ and $B \to X_s \gamma$ data,  model-independent  constraints on the effective couplings of higher-dimensional operators have been derived. While the fit does not provide a unique best solution, a particularly simple scenario of new physics emerges as a possible explanation, that features the following modifications of Wilson coefficients \cite{Descotes-Genon:2013wba}\footnote{See (\ref{operators}) for the  definition of the corresponding operators.}
\beq \label{eq:bestfit}
\Delta C_7^\gamma \sim  0 \,, \qquad \Delta C_9^\ell \sim -1.5 \,, \qquad   \Delta C_{10}^\ell \sim  0 \,.
\eeq
Given that within the SM one has $(C_{9}^\ell)_{\rm SM} \simeq 4.1$ (see~e.g.~\cite{Bobeth:2003at}) the solution~(\ref{eq:bestfit}) should originate from new flavour dynamics that induces a large destructive contribution to the semi-leptonic vector operator, while leaving the electromagnetic dipole and the semi-leptonic axial-vector operator essentially SM-like. 

A glimpse at the extensive literature on quark-flavour physics readily shows that the pattern seen in (\ref{eq:bestfit}) is highly non-standard, and that the usual suspects -- such as the minimal supersymmetric SM \cite{Haisch:2012re,Altmannshofer:2012ks}, warped extra dimension scenarios \cite{Blanke:2008yr,Bauer:2009cf} or models with partial compositeness \cite{KerenZur:2012fr,Straub:2013zca} to just name a few -- cannot accommodate the observed deviations (this point has also been stressed in \cite{Altmannshofer:2013foa, Gauld:2013qba}). An apparent  though ad hoc way to obtain~(\ref{eq:bestfit}) is to postulate the existence of a $Z^\prime$ boson with mass in the TeV range and specific couplings to fermions~\cite{Descotes-Genon:2013wba}: the new neutral gauge boson should couple only  to the left-handed $\bar s b$ current and proportionally to the product $V_{ts}^\ast V_{tb}$ of Cabibbo-Kobayashi-Maskawa~(CKM) matrix elements so that excessive CP-violating contributions to $B_s$--$\bar B_s$ mixing are avoided; the $Z^\prime$  boson should  furthermore couple to  left-handed and right-handed  muons  with  close to equal strength, since the $B \to K^\ast \mu^+ \mu^-$ data seem to prefer a vector rather than an axial-vector coupling to the $\bar \mu \mu$ current.

In this article we discuss a $Z^\prime$-boson model based on the gauge group $SU(3)_c \times SU(3)_L \times U(1)_X$, which has the above properties.  Many variants of these  so-called $3$-$3$-$1$ scenarios have been considered in the literature, but we will focus   on the original proposal~\cite{Pisano:1991ee,Frampton:1992wt} in which the electric charge is given by $Q = T^3 - \sqrt{3} \hspace{0.5mm} T^8 + X$. Here $T^{a} = \lambda^{a}/2$ denotes the $SU(3)_L$ generators with $\lambda^a$ the Gell-Mann matrices and $X$ is the $U(1)_X$ quantum number. This model (for obvious reasons referred to as ``$\beta = -\sqrt{3}\hspace{0.75mm}$'' hereafter) is the only $3$-$3$-$1$ scenario that has the desired feature that the $Z^\prime$ boson couples much more strongly to the vector rather than axial-vector component of the charged-lepton current. As it turns out, the choice $\beta = -\sqrt{3}\hspace{0.75mm}$ also makes the $3$-$3$-$1$ model leptophilic/hadrophobic leading to a rich and interesting phenomenology. 

We find that  correlations between the quark-flavour-changing $b \to s \ell^+\ell^-, \nu \bar \nu$ transitions and modifications in muon decay, quark $\beta$-decay as well as  parity-violating $e^- \to e^-$ observables are unavoidable in the considered $Z^\prime$-boson scenario. Advances in the low-energy measurements of the properties of charged leptons can hence provide valuable insights into the structure of the underlying theory, if  the  deviations in $B \to K^\ast \mu^+ \mu^-$ as seen by LHCb are indeed due to interactions of a new neutral gauge boson. We also stress the importance of  improved lattice-QCD determinations of the hadronic parameters entering the SM prediction for the mass difference in $B_s$--$\bar B_s$ mixing, which provides the strongest constraint on the simplest explanations of the $B \to K^\ast \mu^+ \mu^-$ anomaly in the  $3$-$3$-$1$ model with $\beta = -\sqrt{3}\hspace{0.75mm}$. From the experimental side, updated BaBar and Belle measurements of the decay distributions in $B \to X_s \ell^+ \ell^-$ would be essential to  corroborate or challenge the case for new physics in $|\Delta B| = 1$ transitions. We finally highlight the complementarity between low-$p_T$ and high-$p_T$ measurements by considering the present and future constraints on the $Z^\prime$-boson mass $M_{Z^\prime}$ arising from direct $Z^\prime$-boson searches at the LHC, its  high-luminosity upgrade~(HL-LHC) and an ILC.

Our  work is organised as follows. In Section~\ref{sec:FCNC} we review the generic structure of flavour-changing $Z^\prime$-boson interactions in $3$-$3$-$1$ scenarios and discuss  viable flavour alignments necessary to curb the amount of CP violation in the quark-flavour sector. The explicit expression for the $Z^\prime$-boson couplings relevant for our analysis are given in Section~\ref{sec:couplings}.  Our phenomenological analysis that ranges from the study of $B$-physics observables over precision measurements to direct searches starts with Section \ref{sec:operators} and ends with Section~\ref{sec:direct}. We conclude in Section~\ref{sec:blabla} by summarising our main findings and discussing further ways to cast light on the origin of the deviations seen in the recent $B \to K^\ast \mu^+ \mu^-$ data. 

\section{Flavour-changing quark interactions}
\label{sec:FCNC}

The left-handed couplings between the $Z^\prime$ boson and the SM quarks have in $3$-$3$-$1$ scenarios the generic form 
\beq \label{eq:L}
{\cal L} \supset \sum_{q=d,u} ({\cal{G}}_{q_L})_{ij} \, \bar q_L^i \slashed{Z}^{\hspace{0.5mm} \prime} q_L^j \,.
\eeq
In the weak interaction basis indicated by a superscript ``I'', the couplings are diagonal $3 \times 3$ matrices in flavour space
\beq \label{eq:GammaIs}
{\cal{G}}_{q_L}^I = {\rm diag} \left (g_q^l, g_q^l, g_q^h \right ) \,.
\eeq
Since $g_q^l \neq g_q^h$ the strength of the $Z^\prime$-boson couplings to the third and the first two generations are different. Note that we do not consider the right-handed $Z^\prime$-boson couplings to the SM quarks in this section, since it is always possible to make these interactions flavour diagonal by an appropriate choice of quantum numbers. 

By appropriate unitary rotations $U_{u,d}$ of the chiral quark fields, one can always choose a basis where the mass matrices of the SM quarks are diagonal. In such a basis the $Z^\prime$-boson couplings are given by 
\beq \label{eq:Gammas}
{\cal{G}}_{q_L} = U^\dagger_q \,  {\cal{G}}_{q_L}^I \, U_{q} \,,
\eeq
where the matrices $U_{d,u}$ have to fulfil the constraint
\beq \label{eq:V}
V = U_u^\dagger \, U_d \,,
\eeq
with $V$ denoting the CKM matrix. Notice that in the mass eigenstate basis the matrices in (\ref{eq:Gammas}) will generically contain off-diagonal elements which signals the presence  flavour-changing $Z^\prime$-boson tree-level interactions. 

The flavour-changing quark interactions can however be confined to the sector of down-type quarks by choosing $U_u = 1$ (or equivalently $U_d = V$), i.e.~by alignment in the up-type quark sector. In such a case, one obtains 
\beq \label{eq:upalign} 
{\cal{G}}_{d_L} =  g_q^l \, \bm{1}_3 + \left (  g_q^h -  g_q^l \right ) 
\begin{pmatrix} |V_{td} |^2 & V^\ast_{td} \hspace{0.25mm} V_{ts} &  V^\ast_{td}  \hspace{0.25mm}  V_{tb} \\ 
V^\ast_{ts}  \hspace{0.25mm}  V_{td} & | V_{ts}|^2 &  V^\ast_{ts}  \hspace{0.25mm}  V_{tb} \\ 
V^\ast_{tb}  \hspace{0.25mm}  V_{td} & V^\ast_{tb} V_{ts} &  | V_{tb} |^2 \end{pmatrix}  \,, \qquad 
{\cal{G}}_{u_L} =  {\rm diag} \left (g_q^l, g_q^l, g_q^h \right ) \,.
\eeq
This implies that for  up-type  alignment the $s_L \to d_L Z^\prime$,  $b_L \to d_L Z^\prime$ and $b_L \to s_L Z^\prime$ amplitudes are proportional to the combinations $V_{td}^\ast \hspace{0.25mm} V_{ts}$,  $V_{td}^\ast \hspace{0.25mm} V_{tb}$ and $V_{ts}^\ast \hspace{0.25mm} V_{tb}$  of CKM elements. The flavour-changing $Z^\prime$-boson contributions hence follow the pattern of minimal-flavour violation (MFV). In particular, there will be no new sources of CP violation beyond the CKM phase. 

Given the freedom in the choice of down-type misalignment $U_d$ it is also possible to set the contributions to the $s_L \to d_L Z^\prime$ and   $b_L \to d_L Z^\prime$ to zero. This is achieved by the texture~(see e.g.~\cite{Haisch:2011up})
\beq \label{eq:Ud}
U_d \simeq \begin{pmatrix} 1 & 0&  0\\ 
0&1 &V_{ts}^\ast \\ 
0 & V_{ts} &  V_{tb}\end{pmatrix} \,.
\eeq
It follows that 
\beq \label{eq:GammaPAd}
({\cal{G}}_{d_L})_{23} \simeq \left (g_q^h - g_q^l \right ) V_{ts}^\ast \hspace{0.25mm} V_{tb} \,, \qquad 
({\cal{G}}_{d_L})_{12} \simeq ({\cal{G}}_{d_L})_{13} \simeq  0 \,,
\eeq
to leading power in the Cabibbo angle $\lambda \simeq 0.23$, while the $c_L \to u_L Z^\prime$ transition is governed by the coupling 
\beq \label{eq:GammaPAu}
\begin{split}
({\cal{G}}_{u_L})_{12} & \simeq  \left (g_q^h - g_q^l \right )   \Big [ V_{cb}^\ast \hspace{0.25mm} V_{ub} \, |V_{tb}|^2 +  V_{cb}^\ast \hspace{0.25mm} V_{us} \hspace{0.25mm} V_{ts}^\ast \hspace{0.25mm} V_{tb} +   V_{cs}^\ast \hspace{0.25mm} V_{us} \, |V_{ts}|^2 +   V_{cs}^\ast \hspace{0.25mm} V_{ub}  \hspace{0.25mm} V_{tb}^\ast \hspace{0.25mm} V_{ts}  \Big ] \\[2mm] & \simeq  \left (g_q^h - g_q^l \right ) \,  \frac{A^2 \lambda^7}{2}  \left ( \bar \rho - i \bar \eta -1\right )\,.
\end{split}
\eeq
This result has to be compared with the expression $({\cal{G}}_{u_L})_{12} \simeq  \left (g_q^h - g_q^l \right )  A^2 \lambda^5 \left (\bar \rho - i \bar \eta \right )$ obtained in the case of down-type alignment (i.e.~$U_d = 1$). We see that enforcing flavour-changing effects in the $B_s$-meson sector only leads to new CP violation in the $D$-meson sector with respect to MFV. The resulting effects are however strongly Cabibbo suppressed and we will not consider them any further.  

In the following we will assume that a mechanism is at work that leads to a flavour structure like (\ref{eq:upalign}) or~(\ref{eq:GammaPAd}) suitable to explain the pattern (\ref{eq:bestfit}). In fact, the flavour-changing interactions will necessarily have a non-trivial structure when there are gauge quantum numbers that distinguish generations. This is the case in $Z^\prime$-boson models of the $3$-$3$-$1$ type in which the third-generation fermions are treated differently from  the second and first generation $\big($see~(\ref{eq:GammaIs})$\big)$. The textures of flavour-changing neutral currents will then ultimately be controlled by the symmetry breaking patterns of the horizontal global flavour symmetries. We leave the underlying mechanism and the ultraviolet (UV) origin that gives rise to the flavour alignment  unspecified.

\section{Couplings in $\bm{\beta = -\sqrt{3}}$ model}
\label{sec:couplings}

Assuming up-type alignment (\ref{eq:upalign}), the $Z^\prime$-boson couplings in the $3$-$3$-$1$ model with $\beta = -\sqrt{3}$ relevant for our analysis are $(i \neq j)$
\beq \label{eq:GammadL23} 
 ({\cal{G}}_{d_L})_{ij} = \frac{g \hspace{0.25mm} c_W}{\sqrt{3}  \hspace{0.5mm}  \sqrt{1- 4 s_W^2}} \, V_{ti}^\ast \hspace{0.25mm} V_{tj} \,, \qquad 
 ({\cal{G}}_{u_L})_{ij} =0 \,, \\
\eeq
and (see e.g.~\cite{Buras:2012dp})
\beq \label{eq:diagonal}
\begin{split} 
({\cal{G}}_{d_L})_{ii} & = \frac{g \left ( -1 + 2 c_W^2 \hspace{0.25mm} |V_{ti}|^2 \right )}{2 \sqrt{3} \hspace{0.5mm} c_W \hspace{0.5mm}  \sqrt{1- 4 s_W^2}} \,, \qquad 
({\cal{G}}_{d_R})_{ii}  =  \frac{g  \hspace{0.25mm} s_W^2}{\sqrt{3}  \hspace{0.5mm} c_W   \hspace{0.5mm}\sqrt{1- 4 s_W^2}}  \,,\\[2mm]
({\cal{G}}_{u_L})_{ii} & = \frac{g \left ( -1 + 2 c_W^2 \hspace{0.25mm}\delta_{ti} \right )}{{2 \sqrt{3}  \hspace{0.5mm} c_W   \hspace{0.5mm}\sqrt{1- 4 s_W^2}}} \,, \qquad \hspace{0.05cm}
({\cal{G}}_{u_R})_{ii}  =  \frac{-2 g  \hspace{0.25mm} s_W^2}{{\sqrt{3}  \hspace{0.5mm} c_W   \hspace{0.5mm}\sqrt{1- 4 s_W^2}}} \,,\\[2mm]
({\cal{G}}_{\ell_L})_{ii} & =  \frac{g \left ( 1 + 2 s_W^2 \right)}{2 \sqrt{3}  \hspace{0.5mm} c_W   \hspace{0.5mm}\sqrt{1- 4 s_W^2}} \,, \qquad \hspace{0.75mm}
({\cal{G}}_{\ell_R})_{ii}  =  \frac{\sqrt{3} \, g \, s_W^2}{c_W   \hspace{0.5mm}\sqrt{1- 4 s_W^2}} \,,\\[4mm]
({\cal{G}}_{\nu_L})_{ii}  & = ({\cal{G}}_{\ell_L})_{ii} \,, \qquad \hspace{2.1cm} ({\cal{G}}_{\nu_R})_{ii} = 0 \,, 
\end{split}
\eeq
where $g$ is the usual $SU(2)_L$ coupling, while $s_W$ and $c_W$ denote the sine and cosine of the weak mixing angle. For partial misalignment in the down-type quark with a texture (\ref{eq:GammaPAd}), the coupling $({\cal{G}}_{d_L})_{23}$ is also given by (\ref{eq:GammadL23}), while the remaining off-diagonal entries  $({\cal{G}}_{d_L})_{ij}$  vanish to excellent approximation. The diagonal quark couplings in (\ref{eq:diagonal}) are to first order independent of such a change.  

Notice that for the choice  $\beta = -\sqrt{3}$, the vector and axial-vector couplings of the $Z^\prime$-boson to charge leptons take the form 
\beq \label{eq:GammaVA}
\begin{split}
({\cal{G}}_{\ell_V})_{ii} & = ({\cal{G}}_{\ell_R})_{ii} + ({\cal{G}}_{\ell_L})_{ii} = \frac{g \left ( 1 + 8 s_W^2 \right)}{2 \sqrt{3}  \hspace{0.5mm} c_W   \hspace{0.5mm}\sqrt{1- 4 s_W^2}} \simeq 3.30 \, g \,,   \\[2mm]
({\cal{G}}_{\ell_A})_{ii} & = ({\cal{G}}_{\ell_R})_{ii} - ({\cal{G}}_{\ell_L})_{ii} = -\frac{g \, \sqrt{1 -  4  s_W^2}}{2 \sqrt{3}  \hspace{0.5mm} c_W} \simeq -0.09 \, g \,.
\end{split}
\eeq
Electrons, muons and taus hence couple essentially vectorially to the new gauge boson. For other common choices of $\beta$ (i.e.~$\sqrt{3}$, $\pm 1/\sqrt{3}$) this is not the case, which renders these models uninteresting for our purposes.

\section{Dipole and semi-leptonic operators}
\label{sec:operators}

\begin{figure}[t!]
\begin{center}
\makebox{\includegraphics[width=0.8\columnwidth]{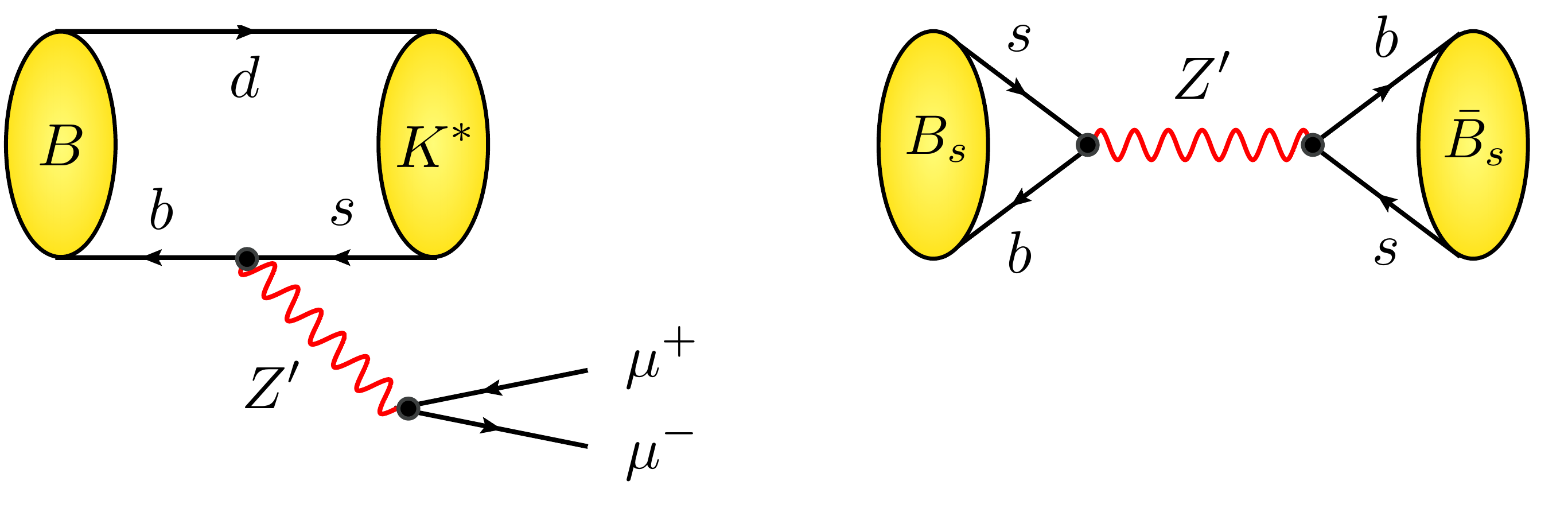}}  
\end{center}
\vspace{-4mm}
\caption{Tree-level contributions to  $B \to K^\ast \mu^+ \mu^-$ (left diagram) and $B_s$--$\bar B_s$ mixing (right diagram) from $Z^\prime$-boson exchange.}
\label{fig:Bphysics}
\end{figure}

In order to calculate the various $b\to s \gamma, \mu^+ \mu^-$ observables in the $3$-$3$-$1$ model, one has to determine the Wilson coefficients of the operators that enter the effective Hamiltonian
\beq
{\cal H}_{\rm eff} = - \frac{4 G_F}{\sqrt{2}} \, V_{ts}^\ast V_{tb} \left ( C_7^\gamma \hspace{0.25mm} Q_7^\gamma + C_9^\ell  \hspace{0.25mm}  Q_9^\ell + C_{10}^\ell  \hspace{0.25mm}  Q_{10}^\ell   \right )  + {\rm h.c.} \,,
\eeq
where $G_F \simeq 1.167 \cdot 10^{-5} \, {\rm GeV}^{-2}$ is the Fermi constant and  ($\ell = e, \mu, \tau$)
\beq \label{operators}
\begin{split}
Q_7^\gamma &=\frac{e}{(4 \pi)^2} \, m_b \left (\bar s_L   \sigma_{\alpha \beta} b_R \right ) F^{\alpha \beta} \,,  \\
Q_9^\ell &= \frac{e^2}{(4 \pi)^2}  \left (\bar s_L   \gamma_\alpha b_L \right )\left ( \bar \ell \gamma^\alpha \ell \right ) \,, \\
Q_{10}^\ell &= \frac{e^2}{(4 \pi)^2}  \left (\bar s_L   \gamma_\alpha b_L \right )\left ( \bar \ell \gamma^\alpha \gamma_5 \ell \right ) \,.
\end{split}
\eeq
A straightforward matching calculation gives
\beq \label{eq:matching}
\begin{split}
\Delta C_7^\gamma & \simeq \frac{8 s_W^2}{27 \left (1 - 4 s_W^2\right )} \, \frac{M_W^2}{M_{Z^\prime}^2} \,, \\[1mm]
\Delta C_9^\ell & =  -\frac{2 \pi}{3 \hspace{0.25mm} \alpha} \, \frac{1 + 8 s_W^2}{1-4s_W^2}  \, \frac{M_W^2}{M_{Z^\prime}^2}  \,, \\[1mm]
\Delta C_{10}^\ell & =  \frac{2 \pi}{3 \hspace{0.25mm} \alpha} \, \frac{M_W^2}{M_{Z^\prime}^2} \,,
\end{split}
\eeq
in agreement with the results presented in \cite{Buras:2012dp}.  To obtain the result for $\Delta C_7^\gamma$ we have only kept  the terms that are not CKM suppressed. The tree-level diagram giving rise to $\Delta C_{9,10}^\ell$ is shown on the left-hand side in Figure~\ref{fig:Bphysics}. Interestingly, our $3$-$3$-$1$ model predicts  $\Delta C_9^\ell < 0$ and $\Delta C_{10}^\ell/\Delta C_9^\ell = -(1-4s_W^2)/(1+8s_W^2) \ll 1$. 

Employing $\alpha= \alpha (M_Z)  \simeq 1/128$,  $s_W^2 = s_W^2 (M_Z) \simeq 0.23$ and $M_W \simeq 80.4 \, {\rm GeV}$, it hence follows that the 68\% confidence level (CL) range 
\beq \label{eq:dc9}
\Delta C_9^\ell \in [-1.9, -1.3] \,,
\eeq
found in \cite{Descotes-Genon:2013wba} from a fit to the present $b\to s \gamma, \mu^+ \mu^-$ data, can be achieved  for $Z^\prime$-boson masses 
\beq \label{eq:MZp}
M_{Z^\prime} \in [5.7, 6.9 ] \; {\rm TeV} \,.
\eeq
Such large $Z^\prime$-boson masses lead to new-physics effects of 
\beq \label{eq:dc7}
\Delta C_7^\gamma = {\cal O} (10^{-4}) \,,
\eeq
i.e.~negligible corrections with respect to $(C_7^\gamma)_{\rm SM} \simeq  -0.19$~\cite{Misiak:2004ew}. Likewise, the $Z^\prime$-boson contributions to the semi-leptonic axial-vector operator are very small, amounting to 
\beq \label{eq:dc10}
\Delta C_{10}^\ell \in [0.04, 0.05] \,.
\eeq
The smallness of the coefficient $\Delta C_7^\gamma$ ($\Delta C_{10}^\ell$) is of course a result of the one-loop suppression of dipole interactions (the vector-like nature of the $Z^\prime$-boson couplings to charged leptons). 

\section{Mass difference in $\bm{B_s}$--$\bm{\bar B_s}$ mixing}
\label{sec:mix}

Tree-level exchange of $Z^\prime$ bosons also leads to a shift in the mass difference of  neutral mesons. From the right graph in Figure~\ref{fig:Bphysics}  we find, in the case of $B_s$--$\bar B_s$ mixing
\beq \label{eq:CBs}
\Delta_{B_s} = \frac{\Delta M_{B_s}}{(\Delta M_{B_s})_{\rm SM}} - 1= \eta_{B_s} \, \frac{16 \pi}{3 \hspace{0.25mm} \alpha} \frac{c_W^2 \hspace{0.25mm} s_W^2}{1- 4 s_W^2} \,\frac{M_W^2}{M_{Z^\prime}^2} \,,
\eeq
with 
\beq \label{eq:etaBs}
\eta_{B_s} \simeq 0.41 \left [ 1 - 0.03 \, \ln \left ( \frac{M_{Z^\prime}}{5 \, {\rm TeV}} \right)\right ] \,.
\eeq
These results agree with  \cite{Buras:2012dp} (also~\cite{Haisch:2011up}). Similar expressions apply in the case of up-type alignment for  $B_d$--$\bar B_d$ and $K$--$\bar K$ mixing. Notice that $\Delta_{B_s}$ is strictly positive, meaning that the $Z^\prime$-boson corrections necessarily enhance the mass difference $\Delta M_{B_s}$ with respect to the SM prediction. This general feature has also been stressed recently in~\cite{Buras:2013qja}. 

Using the available experimental information on $|\Delta F |=2$  processes, the ${\rm UT}_{fit}$ collaboration obtains the following 95\% CL bound \cite{UTFit} 
\beq \label{eq:CBsUTfit}
\Delta_{B_s} \in [-0.16, 0.26] \,.
\eeq
These limits imply that $Z^\prime$-boson masses
\beq \label{eq:MZpBs}
M_{Z^\prime} > 6.9 \, {\rm TeV} \,, 
\eeq
are compatible with the current bounds on $B_s$--$\bar B_s$ mixing. The remaining $|\Delta F |=2$  constraints turn out to be less restrictive and only apply if the flavour structure~(\ref{eq:upalign}) is realised. 

By comparing (\ref{eq:MZpBs}) to (\ref{eq:MZp}) we see that the bound from $B_s$-meson mixing already cuts into the parameter region favoured by $B \to K^\ast \mu^+ \mu^-$. In the $3$-$3$-$1$ model with $\beta = -\sqrt{3}$, the constraint from $\bar B_s$--$B_s$ mixing hence effectively limits the possible new-physics corrections in the Wilson coefficient $C_9^\ell$. Future improvements in lattice-QCD determinations of  the $B_s$-meson decay constant $f_{B_s}$, the hadronic parameter $\hat B_{B_s}$ as well as $V_{cb}$, which represent the dominant sources of uncertainty in $(\Delta M_{B_s})_{\rm SM}$ \cite{Lenz:2006hd}, will therefore be crucial to cross-check $Z^\prime$-boson explanations of the $B \to K^\ast \mu^+ \mu^-$ anomaly. In fact, in our scenario one finds 
\beq \label{eq:monitor}
\Delta C_9^\ell = - \frac{1 + 8 s_W^2}{8 \hspace{0.25mm} c_W^2 s_W^2} \frac{\Delta_{B_s}}{\eta_{B_s}} \simeq -4.9 \hspace{0.25mm} \Delta_{B_s} \,,
\eeq
where in the final step we have neglected the weak logarithmic dependence of $\eta_{B_s}$ on $M_{Z^\prime}$. An upper bound on $\Delta_{B_s}$ hence translates  into a lower bound on $\Delta C_9^\ell$, which makes the relation (\ref{eq:monitor}) useful to monitor  the effect  of future improvements in the SM prediction of  $\Delta M_{B_s}$ on destructive new physics in $B \to K^\ast \mu^+ \mu^-$.

\section{Rare decays: $\bm{B \to X_s \ell^+ \ell^-}$ and $\bm{B \to K \mu^+ \mu^-}$}
\label{sec:rare1}

The large destructive contribution to the semi-leptonic vector operator (\ref{eq:bestfit}) will leave an imprint in both the inclusive $B \to X_s  \ell^+ \ell^-$ decay~($\ell = e, \mu$) as well as the exclusive $B \to K   \mu^+ \mu^-$ channels. For the relative shift in the branching ratio of $B \to X_s  \ell^+ \ell^- $ integrated over low invariant di-lepton masses, i.e.~$q^2 \in [1,6] \, {\rm GeV}^2$, we find by employing the results of \cite{Bobeth:2003at}, the following approximate formula
\beq \label{eq:Bxsll}
\Delta_{X_s}=\frac{{\rm Br}_{\rm \hspace{0.1mm} low} \hspace{-0.5mm} \left (B \to X_s  \ell^+ \ell^- \right )}{{\rm Br}_{\rm \hspace{0.1mm} low} \hspace{-0.5mm}  \left (B \to X_s  \ell^+ \ell^- \right )_{\rm SM}} - 1  \simeq  0.19 \, \Delta C_9^\ell + 0.04 \, \big (  \Delta C_9^\ell \big )^2 \,.
\eeq
Here we have only included the dominant corrections due to $\Delta C_9^\ell$.  Given that $\Delta C_{7}^\gamma \simeq 0$ and $\Delta C_{10}^\ell/\Delta C_9^\ell \simeq -0.03$, the expression (\ref{eq:Bxsll}) is a very good approximation to the full result.  From (\ref{eq:dc9}) it then follows that the  low-$q^2$ branching ratio of $B \to X_s  \ell^+ \ell^- $ should receive a modification of 
\beq \label{eq:DXs}
\Delta_{X_s} \in -[22,18] \hspace{0.5mm}   \%\,,
\eeq
relative to the SM expectation ${\rm Br}_{\rm \hspace{0.1mm} low} \hspace{-0.5mm}  \left (B \to X_s  \ell^+ \ell^- \right )_{\rm SM} = (1.68 \pm 0.17) \cdot 10^{-6}$ \cite{Bauer:2009cf}. 

In this context it is important to emphasise that  the global analyses \cite{Descotes-Genon:2013wba,Altmannshofer:2013foa} both apply the experimental value ${\rm Br}_{\rm \hspace{0.1mm} low} \hspace{-0.5mm}  \left (B \to X_s  \ell^+ \ell^- \right ) = (1.6 \pm 0.5) \cdot 10^{-6}$ obtained in~\cite{Gambino:2004mv} from a naive average of the BaBar \cite{Aubert:2004it} and Belle \cite{Abe:2004sg} measurements. The latter results are by now almost ten years old, but an update of  $B \to X_s  \ell^+ \ell^-$ decay distributions that has been published in a peer-reviewed journal does not exist. There is however a  recent preliminary result from Belle \cite{Bellenew} which is included in the latest HFAG compendium~\cite{Amhis:2012bh} of heavy-flavour averages. It reads ${\rm Br}_{\rm \hspace{0.1mm} low} \hspace{-0.5mm}  \left (B \to X_s  \ell^+ \ell^- \right ) = (0.99 \pm 0.22) \cdot 10^{-6}$ and when averaged with the BaBar measurement~\cite{Aubert:2004it} leads to a value of ${\rm Br}_{\rm \hspace{0.1mm} low} \hspace{-0.5mm}  \left (B \to X_s  \ell^+ \ell^- \right ) = (1.04 \pm 0.21) \cdot 10^{-6}$, which is slightly more than $2 \sigma$ below the SM prediction. Since a shift  $\Delta C_9^\ell \sim -1.5$ results in ${\rm Br}_{\rm \hspace{0.1mm} low} \hspace{-0.5mm}  \left (B \to X_s  \ell^+ \ell^- \right ) \sim 1.3 \cdot 10^{-6}$, the newest Belle results on $B \to X_s  \ell^+ \ell^-$ could also be interpreted as a hint for the presence of a destructive new-physics contribution to the Wilson coefficient of the semi-leptonic vector operator. 

We now turn our attention to the exclusive $b \to s \mu^+ \mu^-$ channels, focusing on the $B^+ \to K^+ \mu^+ \mu^-$ mode which has recently been measured precisely by LHCb \cite{Aaij:2012vr}. In the high-$q^2$ region $[14.18,22] \, {\rm GeV}^2$, we obtain in agreement with \cite{Altmannshofer:2013foa} 
\beq \label{eq:Bkll}
\Delta_{K^+}=\frac{{\rm Br}_{\rm \hspace{0.1mm} high} \hspace{-0.5mm} \left (B^+ \to K^+  \mu^+ \mu^- \right )}{{\rm Br}_{\rm \hspace{0.1mm} high} \hspace{-0.5mm}  \left (B^+ \to K^+  \mu^+ \mu^- \right )_{\rm SM}} - 1  \simeq  0.24 \, \Delta C_9^\ell + 0.03 \, \big (  \Delta C_9^\ell \big )^2 \,, 
\eeq
where $\Delta C_{7}^\gamma$ and $\Delta C_{10}^\ell$ contributions have again been neglected. Using  (\ref{eq:dc9}) it then follows that 
\beq
\Delta_{K^+} \in -[35,26] \hspace{0.5mm}   \%\,.
\eeq
The corresponding SM prediction is ${\rm Br}_{\rm \hspace{0.1mm} high} \hspace{-0.5mm}  \left (B^+ \to K^+  \mu^+ \mu^- \right )_{\rm SM} = (1.10 \pm 0.25) \cdot 10^{-7}$ \cite{Altmannshofer:2013foa}~(see also \cite{Bouchard:2013mia}) and lies on top  of the value ${\rm Br}_{\rm \hspace{0.1mm} high} \hspace{-0.5mm}  \left (B^+ \to K^+  \mu^+ \mu^- \right ) = (1.04 \pm 0.12) \cdot 10^{-7}$ measured by LHCb. A correction $\Delta C_9^\ell \sim -1.5$ leads to ${\rm Br}_{\rm \hspace{0.1mm} high} \hspace{-0.5mm}  \left (B^+ \to K^+  \mu^+ \mu^- \right )\sim 0.8 \cdot 10^{-7}$, corresponding to a tension between theory and experiment of close to $1 \sigma$.    Given the recent observation of a charmonium resonance in the high-$q^2$ di-muon spectrum of $B^+ \to K^+  \mu^+ \mu^-$~\cite{Aaij:2013pta}, 
the theoretical uncertainty plaguing  ${\rm Br}_{\rm \hspace{0.1mm} high} \hspace{-0.5mm}  \left (B^+ \to K^+  \mu^+ \mu^- \right )_{\rm SM}$ deserves further study.

The above discussion makes clear that  the outcome of global analyses of $b \to s$ data and their physics interpretations depend significantly on various factors, most importantly on which  observables are included  in the fit (or excluded from the fit) and the size and treatment of theoretical uncertainties. Keeping this in mind it is not difficult to understand why \cite{Descotes-Genon:2013wba} and \cite{Altmannshofer:2013foa} do not obtain identical results. The main differences in the two analyses are: in the former work the constraint from $B^+ \to K^+  \mu^+ \mu^-$ is not included, while the latter analysis considers this exclusive channel; the global fit in \cite{Descotes-Genon:2013wba} is based on $B \to K^\ast \mu^+ \mu^-$ data from  LHCb alone, whereas  \cite{Altmannshofer:2013foa} incorporates the available results from ATLAS, BaBar, Belle, CDF and CMS; for what concerns the low-$q^2$ region the work  \cite{Descotes-Genon:2013wba} considers the bins $[0.1,2] \, {\rm GeV}^2$,  $[2,4.3] \, {\rm GeV}^2$ and  $[4.3,8.68] \, {\rm GeV}^2$, while the authors of  \cite{Altmannshofer:2013foa} perform averages to obtain values for the observables integrated over the $[1,6] \, {\rm GeV}^2$ bin. These differences in combination with the fact that the most significant deviation of $3.7 \sigma$ appears in  $P_5^\prime$ in the interval $[4.3,8.68] \, {\rm GeV}^2$  \cite{Serra,Aaij:2013qta} explains why, in contrast to \cite{Descotes-Genon:2013wba}, the article \cite{Altmannshofer:2013foa} finds that a new-physics contribution $\Delta C_9^\ell \sim -1.5$ alone does not lead to a good description of the data.

We argued above that also the treatment of  $B \to X_s \ell^+ \ell^-$ in the global fit is likely to have an important impact on the physics implications of the $B \to K^\ast \mu^+ \mu^-$ anomaly (and may in fact also point to a sizeable negative contribution $\Delta C_9^\ell$). Since the theoretical predictions for $B \to X_s \ell^+ \ell^-$ are rather sound,  BaBar and Belle measurements of the inclusive rare semi-leptonic $B$ decay based on the full data sets are urgently needed to clarify the situation.  We also add that the preference for a small negative new-physics contribution to the Wilson coefficient $C_7^\gamma$, as found in both analyses \cite{Descotes-Genon:2013wba, Altmannshofer:2013foa}, is a simple consequence \cite{Haisch:2007ia} of the fact that the SM value of the branching ratio of $B \to X_s \gamma$ \cite{Misiak:2006zs,Misiak:2006ab} is around $1 \sigma$ below the experimental world average \cite{Amhis:2012bh}.

\section{Rare decays: $\bm{B_s \to \mu^+ \mu^-}$ and $\bm{B \to K^{(\ast)}, X_s \nu \bar \nu}$}
\label{sec:rare2}

Besides $B \to K^{(\ast)} \mu^+ \mu^-$  and $B \to X_s \ell^+ \ell^-$ the LHC measurements of $B_s \to \mu^+ \mu^-$~\cite{Aaij:2013aka,Chatrchyan:2013bka} also  put tight constraints on the Wilson coefficients of the semi-leptonic operators. In fact, only the axial-vector operator  enters the prediction for the  $B_s \to \mu^+ \mu^-$ rate
\beq \label{eq:Rmm}
\Delta_{\mu^+  \mu^-} = \frac{{\rm Br} \hspace{-0.5mm} \left (B_s \to \mu^+ \mu^- \right )}{{\rm Br} \hspace{-0.5mm}  \left (B_s \to \mu^+ \mu^- \right )_{\rm SM}} - 1 = \frac{\big |(C_{10}^\ell)_{\rm SM} + \Delta C_{10}^\ell \big |^2}{\big |(C_{10}^\ell)_{\rm SM} \big |^2}  - 1 \,.
\eeq
Recalling that $(C_{10}^\ell)_{\rm SM} \simeq -4.3$ (see e.g.~\cite{Bobeth:2003at}), the range (\ref{eq:dc10}) translates into 
\beq \label{eq:Rmmrange}
\Delta_{\mu^+  \mu^-} \in -[2.5, 1.7] \hspace{0.5mm}  \% \,,
\eeq
corresponding to suppressions of the branching fraction by ${\cal O} (2\%)$ relative to the SM. Such small effects are (and will remain)  unobservable given the theoretical uncertainties inherent in $B_s \to \mu^+ \mu^-$. Similarly, the decay modes $B_d  \to \mu^+ \mu^-$ and $K_L \to \pi^0 \mu^+ \mu^-$ will receive only very small or no corrections in our $3$-$3$-$1$ model depending on whether the flavour structure~(\ref{eq:upalign}) or (\ref{eq:GammaPAd}) is realised. 

In the future the rare decays $B \to K^{(\ast)}, X_s \nu \bar \nu$  may also allow for transparent studies of electroweak penguin effects. Normalised to the SM rates, one finds for the branching ratios of these modes 
\beq \label{eq:Rnn}
\Delta_{\nu \bar \nu} = \frac{{\rm Br} \hspace{-0.5mm} \left (B \to K^{(\ast)}, X_s \nu \bar \nu \right )}{{\rm Br} \hspace{-0.5mm}  \left (B \to  K^{(\ast)}, X_s  \nu \bar \nu \right )_{\rm SM}} - 1 = \frac{\big |X_{\rm SM}+ \Delta X \big |^2}{\big |X_{\rm SM} \big |^2} - 1 \,,
\eeq
with $X_{\rm SM} \simeq 1.47$ \cite{Brod:2010hi}. The coefficient $\Delta X$ takes the form (cf.~also \cite{Buras:2012dp})
\beq \label{eq:dx}
\Delta X = \frac{2\pi}{3 \hspace{0.25mm} \alpha} \, \frac{s_W^2 \left ( 1 + 2 s_W^2 \right )}{1-4s_W^2}  \, \frac{M_W^2}{M_{Z^\prime}^2}  \,.
\eeq
For $Z^\prime$-boson masses in the range (\ref{eq:MZp}), one obtains
\beq
\Delta_{\nu \bar \nu} \in  [22, 33] \hspace{0.5mm} \% \,,
\eeq
implying enhancements of the $b \to s \nu \bar \nu$ branching ratios of ${\cal O} (25\%)$. Note that this is a rather model-independent conclusion following from $SU(2)_L$ invariance, lepton-flavour universality and the absence/smallness of right-handed currents \cite{Altmannshofer:2013foa,Gauld:2013qba,Altmannshofer:2009ma}. If the up-type sector is fully aligned, effects of the same size arise in $B \to X_d \nu \bar \nu$, while the corrections in $K \to \pi \nu \bar \nu$  are smaller and follow the MFV pattern, i.e.~$\Delta {\rm Br} \hspace{-0.5mm} \left (K^+ \to \pi^+ \nu \bar \nu \right )/\Delta {\rm Br} \hspace{-0.5mm} \left (K_L \to \pi^0 \nu \bar \nu \right ) \simeq 2.2$. Misaligning the down-type sector will break the correlations between $b \to s \nu \bar \nu$ and the $b \to d \nu \bar \nu$, $s \to d \nu \bar \nu$ transitions. 

\section{Non-leptonic decays}
\label{sec:nonlep}

As another application, we consider the puzzle of the difference $\Delta A_{\rm CP} = (12.6 \pm 2.2) \%$~\cite{Hofer:2012vc} in the direct CP asymmetries $A_{\rm CP} (B^- \to \pi^0 K^-)$ and $A_{\rm CP} (B^0 \to \pi^+ K^-)$. In the SM, theoretical expectations for  $\Delta A_{\rm CP}$ are typically no more than a few percent $\big($e.g.~\cite{Bauer:2009cf} quotes $ ( \Delta A_{\rm CP} )_{\rm SM} = (0.7 \pm 2.9) \%$$\big)$. Within the $3$-$3$-$1$ model under consideration, the non-leptonic $B$-meson decays receive corrections from the effective Hamiltonian 
\beq
{\cal H}_{\rm eff} = - \frac{4 G_F}{\sqrt{2}} \, V_{ts}^\ast V_{tb} \sum_{i=3,5,7,9} C_i  \hspace{0.25mm} Q_i  + {\rm h.c.} \,,
\eeq
which contains the usual QCD and electroweak penguin operators\footnote{As the couplings $({\cal{G}}_{d_L})_{ii}$ in (\ref{eq:diagonal}) are flavour non-universal also the operator $Q_1^b = (\bar s_L \gamma_\alpha b_L) (\bar b_L \gamma^\alpha b_L)$~\cite{Bobeth:2003at} will get a non-zero initial condition. We neglect this effect since its impact on the processes under consideration is very small.}
\beq \label{eq:QCDEW}
\begin{split}
Q_3 & =  \left (\bar s_L \gamma_\alpha b_L \right) \sum_{q=u,d,s,c,b}  \left (\bar q_L \gamma^\alpha q_L \right) \,, \qquad
\hspace{0.4cm} Q_5 =  \left (\bar s_L \gamma_\alpha b_L \right) \sum_{q=u,d,s,c,b}   \left (\bar q_R \gamma^\alpha q_R \right) \,, \\[1mm]
Q_7& = \frac{3}{2} \left (\bar s_L \gamma_\alpha b_L \right) \sum_{q=u,d,s,c,b}  e_q \left (\bar q_R \gamma^\alpha q_R \right) \,, \quad 
Q_9 = \frac{3}{2} \left (\bar s_L \gamma_\alpha b_L \right) \sum_{q=u,d,s,c,b}    e_q \left (\bar q_L \gamma^\alpha q_L \right) \,, 
\end{split}
\eeq
with $e_q$ denoting the electric charge of the quark $q$. Tree-level $Z^\prime$-boson exchange leads to the following matching corrections  
\beq \label{eq:QCDEQmatching}
\Delta C_3 = \frac{1 }{3 \left ( 1- 4 s_W^2 \right )} \, \frac{M_W^2}{M_{Z^\prime}^2} \,, \qquad
\Delta C_7 =  \frac{4 s_W^2}{3 \left ( 1- 4 s_W^2 \right )}  \, \frac{M_W^2}{M_{Z^\prime}^2}  \,,
\eeq
$\Delta C_5 = 0$ and $\Delta C_9 = 0$.  Employing the results of \cite{Bauer:2009cf}, we find
\beq \label{eq:BKpi}
\Delta A_{\rm CP} - ( \Delta A_{\rm CP} )_{\rm SM}  \simeq 3.3   \left [ 1 + 0.07  \, \ln \left ( \frac{M_{Z^\prime}}{5 \, {\rm TeV}} \right ) \right ]   \Delta C_7 \simeq 13 \; \frac{M_W^2}{M_{Z^\prime}^2}  \,,
\eeq
where the tiny effects due to $\Delta C_3$ have been neglected. The logarithmic dependence on the $Z^\prime$-boson mass arises from renormalisation group  running between the scales $M_{Z^\prime}$ and $M_W$. For $M_{Z^\prime} = {\cal O} (5 \, {\rm TeV})$ one arrives at  $\Delta A_{\rm CP} - ( \Delta A_{\rm CP} )_{\rm SM} = {\cal O} (0.3\%)$, i.e.~the $Z^\prime$-boson corrections in $\Delta A_{\rm CP}$  are clearly insufficient to account for the ${\cal O} (10\%)$ discrepancy in $B \to \pi K$. 

The parameter $\epsilon^\prime/\epsilon$ that measures direct CP violation in $K \to \pi \pi$ is also a very sensitive probe of modifications in the electroweak penguin sector.  Assuming complete up-type alignment in  our $3$-$3$-$1$ model,  we have to excellent approximation (see e.g.~\cite{Bauer:2009cf})  
\beq
\Delta_{\epsilon^\prime/\epsilon} = \frac{\epsilon^\prime/\epsilon}{(\epsilon^\prime/\epsilon)_{\rm SM}} -1 \simeq -2479 \left [ 1 + 0.1  \, \ln \left ( \frac{M_{Z^\prime}}{5 \, {\rm TeV}} \right ) \right ] \Delta C_7 \simeq  -9504 \; \frac{M_W^2}{M_{Z^\prime}^2} \,.
\eeq
Here we have employed the vacuum-insertion approximation  $B_{6,8} =1$ for the hadronic parameters and in the final step neglected the weak logarithmic dependence.  We see that even for $M_{Z^\prime} =  {\cal O} (5 \, {\rm TeV})$ the numerical shifts  $\Delta_{\epsilon^\prime/\epsilon}$  reach (or exceed) the level of $-200\%$. Given the expected progress  in the lattice calculations of $\epsilon^\prime/\epsilon$ such large effects should become clearly visible in the near future.  Since the effects in $b \to s q \bar q$ and $s \to d q \bar q$ can be decoupled by down-type misalignment, an order of magnitude  suppression of $\epsilon^\prime/\epsilon$ is however clearly not a model-independent prediction in the new-physics scenario under study. 

\section{Precision measurements}
\label{sec:PM}

Until now we have only considered the impact of $Z^\prime$-boson exchange in quark-flavour physics. Given that the flavour-diagonal $Z^\prime$-boson  couplings to both quarks and leptons are large  and non-universal~$\big($cf.~(\ref{eq:diagonal})$\big)$, current precision measurements already impose lower limits on $M_{Z^\prime}$. In the following we will neglect effects due to $Z$--$Z^\prime$ mixing that are known to be very tightly bounded (see e.g.~\cite{Langacker:2000ju}). 

We first discuss the modifications $\Delta g_{L\ell}$, $\Delta g_{R\ell}$  of the couplings $g_{L\ell}$,  $g_{R\ell}$ of the $Z$ boson to charged leptons. In our $3$-$3$-$1$ model this is a one-loop effect that in the limit of $M_Z^\prime \gg M_W$ gives rise to a shift \cite{Haisch:2011up} 
\beq
\frac{\Delta g_{L\ell}}{g_{L\ell}} \simeq  \; 0.01 \, \frac{M_W^2}{M_{Z^\prime}^2} \left [ 1 - 0.6 \, \ln \left ( \frac{M_W^2}{M_{Z^\prime}^2}  \right ) \right ] \,,
\eeq
in the left-handed coupling of the charged leptons to the $Z$ boson. The experimental constraint $|\Delta g_{L\ell}|/|g_{L\ell}| \lesssim 1 \permil$ \cite{ALEPH:2005ab} is hence safely fulfilled for $M_{Z^\prime} > 0.5 \, {\rm TeV}$. The constraints that arise from the right-handed charged-lepton couplings as well as the total width $\Gamma_Z$ of the $Z$ boson and the  hadronic pole cross-section $\sigma_{\rm had}$ are similar in strength.  

Bounds from atomic parity violation and polarised electron-electron M\o{}ller scattering asymmetries are also known to provide powerful constraints on heavy $Z^\prime$ bosons if they violate parity (which is the case in our $3$-$3$-$1$ model). For example, the effect due to a $Z^\prime$ boson on the weak charge of a nucleus of $Z$ protons and $N$ neutrons can be written as \cite{Bouchiat:2004sp}
\beq \label{eq:dQW} 
\Delta Q_W (Z, N) = \frac{\sqrt{2}}{2 G_F} \, \frac{({\cal {G}}_{\ell_A})_{11} \, {\cal {G}}_{q_V} (Z,N)}{M_{Z^\prime}^2} \, 3 \left ( Z + N \right ) \,,
\eeq
where
\beq \label{GVqZN}
 {\cal {G}}_{q_V} (Z,N) = \frac{\left ( 2 Z + N \right ) ({\cal {G}}_{u_V})_{11} + \left (  Z + 2 N \right ) ({\cal {G}}_{d_V})_{11} }{3  \left ( Z + N \right )} \,.
\eeq
Numerically, one finds in the case of  Cesium ($^{133}_{\phantom{x}55} {\rm Cs}$)
\beq \label{eq:QWMformula}
\Delta Q_W^{\rm Cs} = \Delta Q_W (55, 78) \simeq 206 \; \frac{M_W^2}{M_{Z^\prime}^2}\,,
\eeq
which combined with the 90\% CL limit $\big |\Delta Q_W^{\rm Cs} \big | <0.6$ \cite{Davoudiasl:2012qa} implies $M_{Z^\prime} > 1.5 \, {\rm TeV}$.
Notice that this is  a non-trivial bound, although the new-physics effect $\Delta Q_W^{\rm Cs}$ is suppressed in our model by the small axial-vector coupling of the electron, $({\cal {G}}_{\ell_A})_{11} \simeq -0.06$.

Similarly to (\ref{eq:dQW}), shifts from the expected  electron-electron M\o{}ller scattering asymmetry can effectively be encoded in the weak charge of the electron 
\beq \label{eq:dQWe} 
\Delta Q_W^e = \frac{\sqrt{2}}{2 G_F} \, \frac{({\cal {G}}_{\ell_A})_{11} \, ({\cal {G}}_{\ell_V})_{11}}{M_{Z^\prime}^2}  \simeq -1.2 \; \frac{M_W^2}{M_{Z^\prime}^2} \,,
\eeq
where the final result corresponds to our $3$-$3$-$1$ model with $\beta = -\sqrt{3}$. The 90\% CL limit $\big |\Delta Q_W^{e} \big | < 0.016$ \cite{Beringer:1900zz} is satisfied for $M_{Z^\prime} > 0.7 \, {\rm TeV}$, meaning that  at present the weak charge of the electron does not pose a very strong bound on the considered scenario. 

\begin{figure}[t!]
\begin{center}
\makebox{\includegraphics[width=0.575\columnwidth]{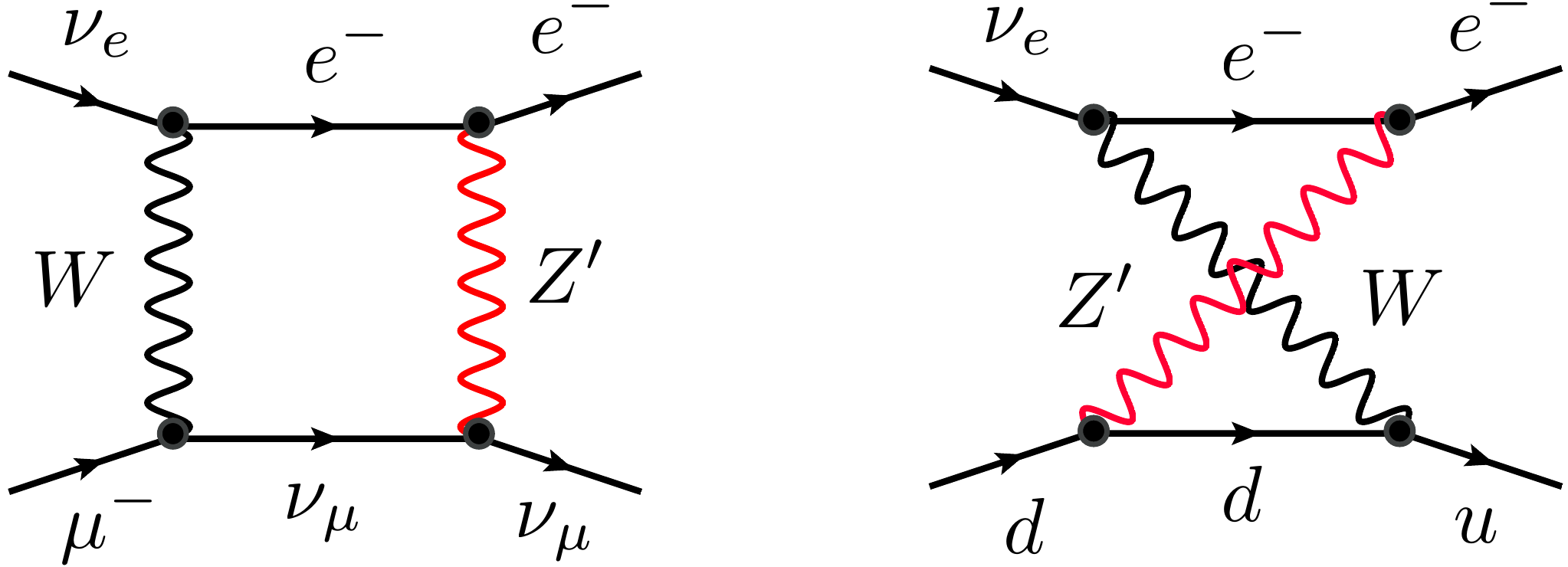}}  
\end{center}
\vspace{-2mm}
\caption{Examples of one-loop box corrections to muon (left) and down-quark (right) decays involving $W$ and $Z^\prime$ bosons.  In the case of $V_{us}$ ($V_{ub}$), the $d$ quark in the right graph has to be replaced by a $s$ ($b$) quark.}
\label{fig:CKM}
\end{figure}

Constraints on additional neutral gauge bosons also derive from violations of first row CKM unitarity as parametrised by
\beq \label{eq:D1}
\Delta_{\rm CKM} = 1 - \sum_{q=d,s,b} |V_{uq}|^2 \,.
\eeq	
The coefficient $\Delta_{\rm CKM}$ is determined from  the difference of the one-loop $Z^\prime$-boson corrections to quark $\beta$-decay amplitudes from which the CKM elements  are extracted as well as muon decay which normalises those amplitudes. The corresponding Feynman diagrams are depicted in Figure~\ref{fig:CKM}. Generalising the classic results of \cite{Marciano:1987ja}, we find the following expression\footnote{Only the logarithmically enhanced terms of the one-loop box corrections to muon and quark decay have been included here. Furthermore, subleading effects involving CKM elements in (\ref{eq:diagonal}) have been neglected.}
\beq \label{eq:D1} 
\Delta_{\rm CKM}  \simeq -\frac{3}{4\pi^2} \, \frac{M_W^2}{M_{Z^\prime}^2} \, \ln \left ( \frac{M_W^2}{M_{Z^\prime}^2} \right )  ({\cal{G}}_{\ell_L})_{11} \, \left[ ({\cal{G}}_{\ell_L})_{11} - \frac{({\cal{G}}_{d_L})_{11} + ({\cal{G}}_{u_L})_{11} }{2}  \right ] \,.
\eeq	
Since  the $SU(2)_L$  relation $({\cal{G}}_{d_L})_{11} =  ({\cal{G}}_{u_L})_{11}$ is broken by small mixing angles only, our expression for $\Delta_{\rm CKM}$ agrees  with the result presented recently in \cite{Buras:2013qja}. Inserting the couplings~(\ref{eq:diagonal}) into the above formula gives  
\beq 
\Delta_{\rm CKM}  \simeq  -0.16 \, \frac{M_W^2}{M_{Z^\prime}^2}  \, \ln \left ( \frac{M_W^2}{M_{Z^\prime}^2} \right )  \,.
\eeq
Employing the experimental bound $|\Delta_{\rm CKM}| < 1\permil$ \cite{Antonelli:2010yf} it then follows that
\beq
M_{Z^\prime} > 2.7 \, {\rm TeV} \,,
\eeq  
at 90\% CL. Notice that while the individual constraints from (\ref{eq:dQW}), (\ref{eq:dQWe}) and (\ref{eq:D1}) can be evaded by suitable choices of the lepton and quark couplings, dodging all bounds at once is not possible.  This shows the complementarity of present and upcoming $\Delta Q_W^{{\rm Cs},p,e}$ and $\Delta_{\rm CKM}$ precision measurements (see Section~\ref{sec:blabla}) in extracting information on  $Z^\prime$-boson models. 

We add finally that the presence of a new neutral gauge boson with close to vector-like couplings to muons leads to a positive shift in the muon anomalous magnetic moment (see e.g.~\cite{Jegerlehner:2009ry}). In the $3$-$3$-$1$ scenario with $\beta = -\sqrt{3}$, one finds $\Delta a_\mu \simeq 2.7 \cdot 10^{-9} \; (0.2 \, {\rm TeV}/M_{Z^\prime})^2$. Clearly, for $M_{Z^\prime}$ values in the TeV range no improvement of the infamous tension between the experimental result and the SM prediction for $a_\mu$ can be achieved. 

\section{Direct searches}
\label{sec:direct}

\begin{figure}
\begin{center}
\makebox{\includegraphics[width=0.8\columnwidth]{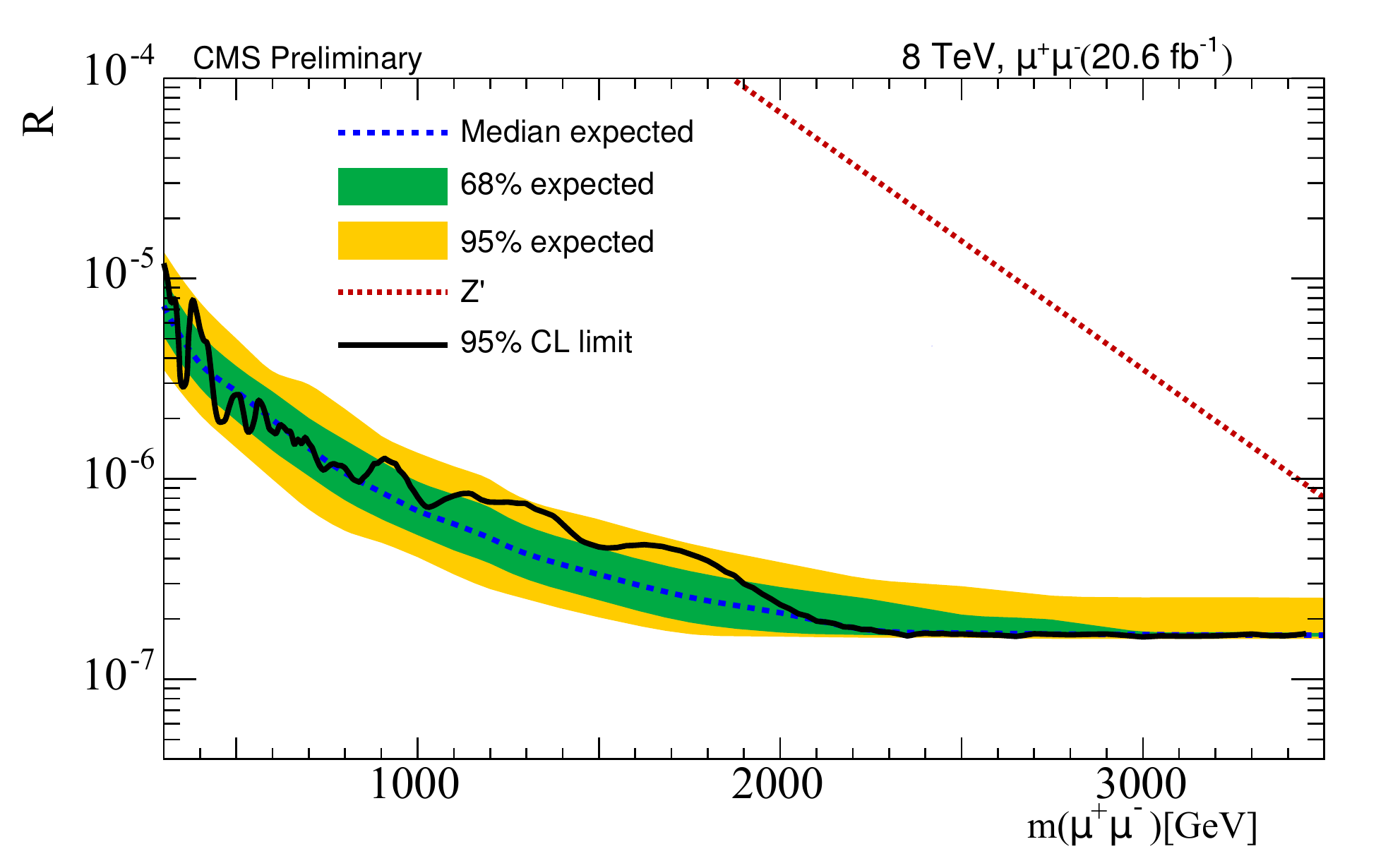}} 
\vspace{0mm}
\caption{Upper limits as a function of $Z^\prime$-boson mass on the production ratio $R$ of cross section times branching fraction into muon pairs. The green and yellow band corresponds to the 68\% and 95\% CLs for the expected limits obtained in  \cite{CMS:2012exa}. The prediction in our $3$-$3$-$1$ model is indicated by the dotted red line. }
\label{directpresent}
\end{center}
\end{figure}

In order to determine the best direct search strategy for  a $Z^\prime$ boson,  one has to consider the partial $Z^\prime$-boson decay widths. From (\ref{eq:diagonal})  we find  that the ratio of charged leptonic to hadronic decays is given in the considered $3$-$3$-$1$ model by 
\beq
\frac{\Gamma  \hspace{-0.5mm} \left (Z^\prime \to \ell^+ \ell^- \right )}{\Gamma  \hspace{-0.5mm}  \left (Z^\prime \to {\rm hadrons} \right )} \simeq 0.52 \,,
\eeq
which is a factor of around 3 larger than what would be obtained if the $Z^\prime$ boson would couple with universal strength to all fermion chiralities. This implies that the most promising channel for an LHC discovery is provided by Drell-Yan~(DY) $Z^\prime$-boson production followed by same-flavour di-lepton decays. Given the significantly larger QCD backgrounds, searches for di-jet resonances and the measurements of the di-jet angular distributions have less potential and thus will not be examined in what follows.

The latest  resonance searches in the di-lepton invariant mass spectrum by ATLAS~\cite{ATLAS:2013jma} and CMS~\cite{CMS:2012exa} both include approximately $20 \, {\rm fb}^{-1}$ of $\sqrt{s} = 8 \, {\rm TeV}$ data. As illustrated in Figure~\ref{directpresent}, the most recent CMS search in the di-muon channel allows to derive a 95\% CL bound of 
\beq
M_{Z^\prime} > 3.9 \,{\rm TeV} \,.
\eeq
In order to derive this number we have implemented the $3$-$3$-$1$ model  into MadGraph5~\cite{Alwall:2011uj} using CTEQ6l1 parton distribution functions~\cite{Pumplin:2002vw} for the event generation while imposing the relevant CMS cuts.  We then extrapolated the expected CMS limit on $R = \sigma (pp \to Z^\prime + X \to \mu^+ \mu^- + X)/ \sigma (pp \to Z + X \to \mu^+ \mu^- + X)$ linearly beyond $3.5 \, {\rm TeV}$ to set the bound. Slightly weaker exclusions are obtained from the di-electron channel of CMS as well as the ATLAS searches. 

\begin{figure}
\begin{center}
\makebox{\includegraphics[width=0.7\columnwidth]{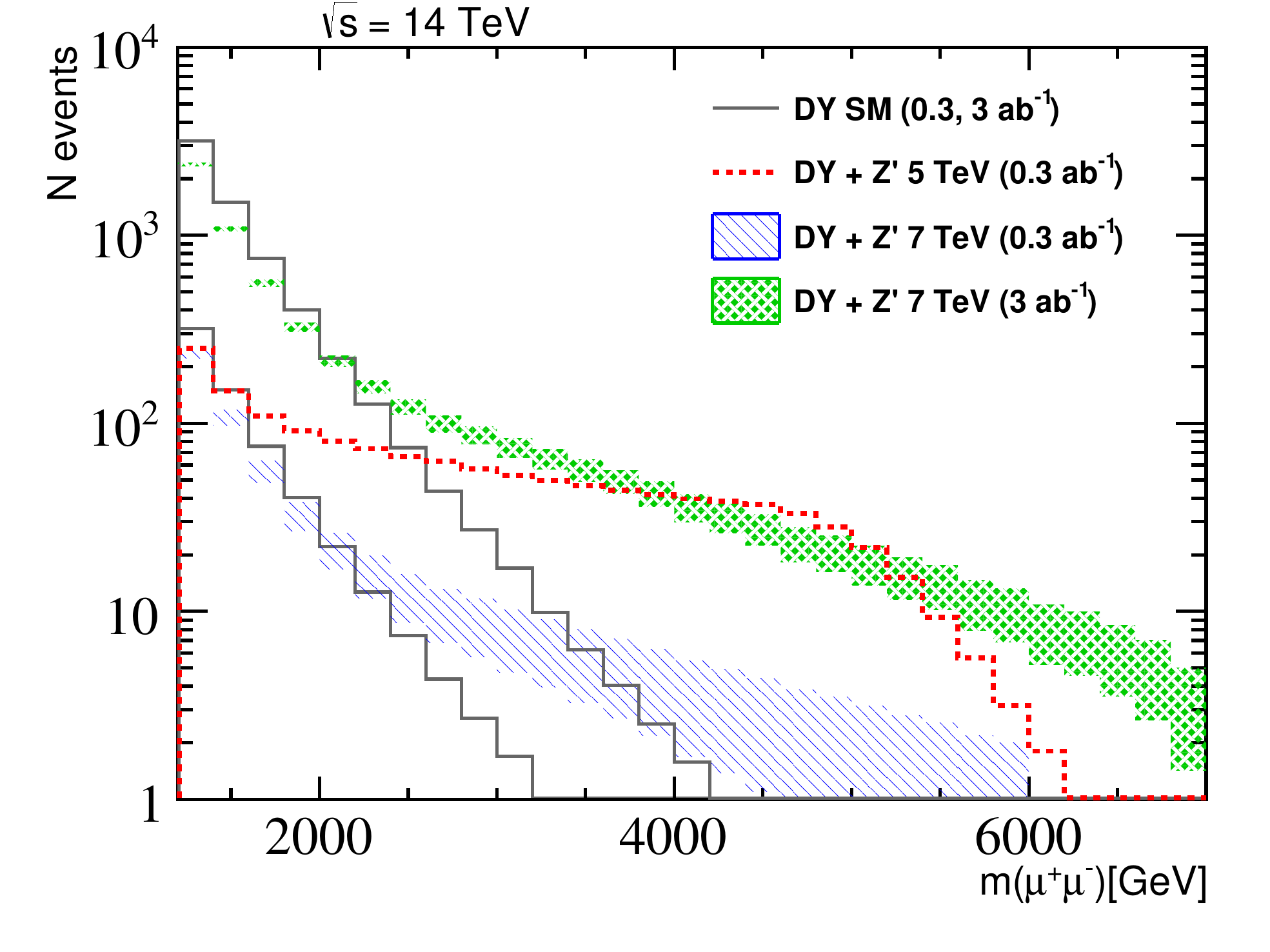}} 
\vspace{-2mm}
\caption{Expected event yield in the invariant mass distribution of muon pairs from DY production at the LHC with $\sqrt{s} = 14 \, {\rm TeV}$, assuming 0.3~ab$^{-1}$   and 3~ab$^{-1}$ of integrated luminosity. The SM expectation and deviations arising from the specific $3$-$3$-$1$ model are plotted for varying $Z^\prime$-boson mass $M_{Z^\prime}$ as indicated in the legend. See text for further explanations.}
\label{directfuture}
\end{center}
\end{figure}

The above limit makes clear that direct $Z^\prime$-boson searches can at present not probe the mass range of ${\cal O}  ( 7 \, {\rm TeV})$ favoured by the $B \to K^\ast \mu^+ \mu^-$ anomaly. To estimate the future LHC prospects we study the expected kinematic and statistical reach in  $pp \to Z^\prime \to \mu^+ \mu^-$ at $\sqrt{s} = 14 \, {\rm TeV}$, assuming integrated luminosities of $0.3$~ab$^{-1}$ and $3 \, {\rm ab}^{-1}$.  The muons are required to be within $|\eta| <$ 2.1 and that at least one of the muons has $p_T > 45 \, {\rm GeV}$.  Our results are shown in Figure~\ref{directfuture}. The dotted red line corresponds to the expected event yield at $0.3 \, {\rm ab}^{-1}$ for a $Z^\prime$-boson with mass of $5 \, {\rm TeV}$, while the blue hatched band indicates the prediction for $M_{Z^\prime} = 7 \, {\rm TeV}$, including statistical errors, assuming the same amount of integrated luminosity.  We see that a $Z^\prime$ boson of $5 \, {\rm TeV}$ should be clearly visible with $0.3 \, {\rm ab}^{-1}$ of $\sqrt{s} = 14 \, {\rm TeV}$ data. However, in the case of $M_{Z^\prime} = 7 \, {\rm TeV}$, we expect only a small excess of events in the tail beyond $m(\mu^+\mu^-) > 3 \, {\rm TeV}$ where the statistics are low. To conclusively probe $Z^\prime$-boson masses up to $7 \, {\rm TeV}$ in the considered $3$-$3$-$1$ model the full HL-LHC data set of $3 \, {\rm ab}^{-1}$ will be necessary -- illustrated by the green hatched band in the plot. Of course, our results should be taken with a grain of salt, since our calculations are simple minded as they are performed at the leading order and do not include the effects of parton showering and hadronisation nor detector response. Yet, we are optimistic that in the  $3$-$3$-$1$ model with $\beta = -\sqrt{3}$, a discovery of a  $Z^\prime$-boson with mass of  ${\cal O} (7 \  {\rm TeV})$, though challenging, should be ultimately possible at the HL-LHC. 

\begin{figure}
\begin{center}
\makebox{\includegraphics[width=0.7\columnwidth]{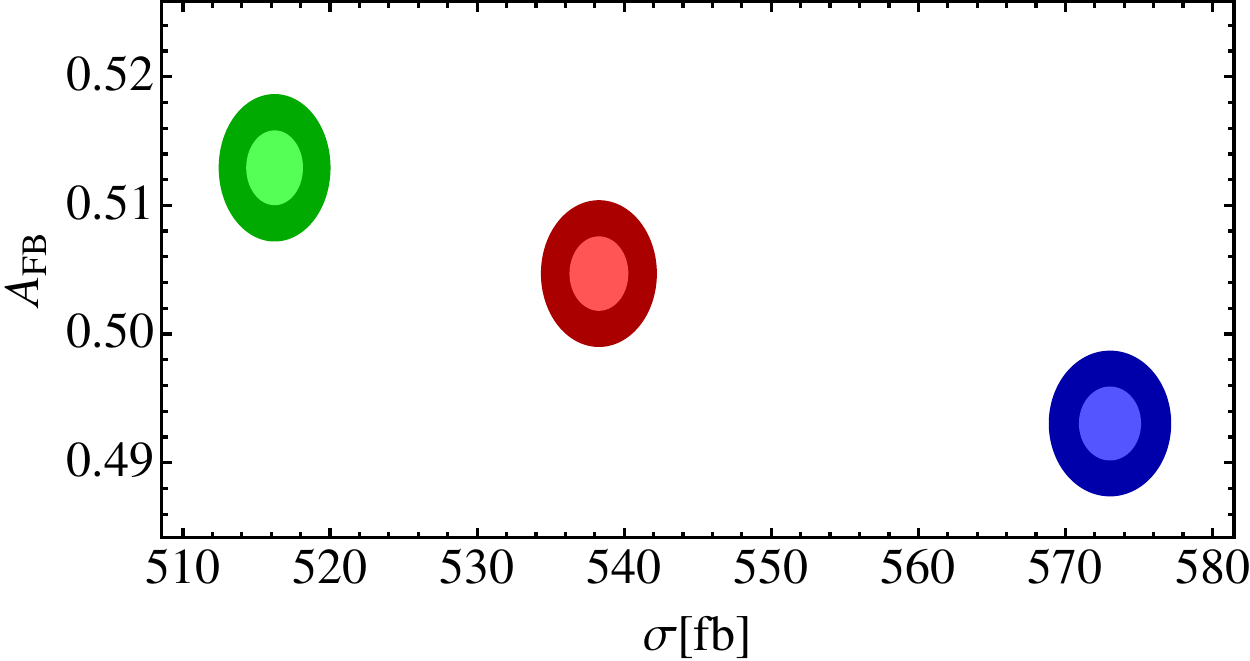}} 
\vspace{1mm}
\caption{Predictions for the $e^+ e^- \to \mu^+ \mu^-$ cross section $\sigma$ and the forward-backward asymmetry~$A_{\rm FB}$ in the SM (blue) and assuming a $Z^\prime$ boson of mass $7 \, {\rm TeV}$ (green) and $9 \, {\rm TeV}$ (red). The light (dark) coloured regions represent 68\% (95\%) CL contours. }
\label{ilc}
\end{center}
\end{figure}

Since a $Z^\prime$ boson interferes with the photon and the $Z$ boson, its couplings to fermions can also be probed  at a high-luminosity $e^+ e^-$ collider. Below we study the mass reach in our $3$-$3$-$1$ scenario of an ILC with $\sqrt{s} = 500 \, {\rm GeV}$ CM energy and beam polarisations  $(P_{e^+}, P_{e^-}) =  (0.3, 0.8)$. For simplicity  we consider only the di-muon final state, which we simulate using MadGraph5. Following \cite{Han:2013mra}  only muons with a polar angle  $\theta \in [10^\circ, 170^\circ]$ are accepted, and we  assume a  total beam polarisation uncertainty of~0.25\% and an intrinsic error of 0.20\% (0.06\%) associated to the measurement of symmetric (asymmetric) leptonic observables. Our results are displayed in Figure~\ref{ilc}. We see that in the presence of a $Z^\prime$~boson with $M_Z^\prime = 7 \, {\rm TeV}$ ($M_Z^\prime = 9 \, {\rm TeV}$) the predicted di-muon cross sections $\sigma$ are smaller than the SM expectation by  $9.9 \%$ ($6.1 \%$). In view of the high-precision measurements possible at an ILC, a reduction of the $e^+ e^- \to \mu^+ \mu^-$ cross section by a few percent should be clearly visible. In the context of the  $3$-$3$-$1$ scenario with $\beta = - \sqrt{3}$ this implies that an $500 \, {\rm GeV}$ $e^+ e^-$ machine can probe  $Z^\prime$-boson masses up to ${\cal O} (10 \, {\rm TeV})$. Further information on the $Z^\prime$-boson couplings to leptons can be obtained by measuring the forward-backward asymmetry $A_{\rm FB}$. It is evident from the figure that in the model under consideration the predicted deviations in $A_{\rm FB}$ are by a factor of  around  3 smaller than those in $\sigma$. Numerically, we find that a $7 \, {\rm TeV}$ ($9 \, {\rm TeV}$) $Z^\prime$~boson leads to a shift of $+4.0\%$~($+2.4\%$)  in $A_{\rm FB}$ relative to the SM.  In contrast to many other  commonly studied $Z^\prime$-boson scenarios~(see~e.g.~\cite{Han:2013mra}) the forward-backward asymmetry is however enhanced in our $3$-$3$-$1$ model relative to the SM prediction.
 
\section{Discussion}
\label{sec:blabla}

Motivated by the recent LHCb results \cite{Serra,Aaij:2013qta} that show a discrepancy of $3.7 \sigma$ in one of the angular observables in $B \to K^\ast \mu^+ \mu^-$, we have studied in this article the $Z^\prime$-boson contributions to various low-$p_T$ and high-$p_T$ observables that arise in the $3$-$3$-$1$ model with $\beta = -\sqrt{3}$. We found that in this specific model, assuming a suitable flavour alignment in the down-type quark sector,  a $Z^\prime$ boson with mass of ${\cal O} (7 \, {\rm TeV})$ leads to a negative shift  in the Wilson coefficient of the semi-leptonic vector operator, which could cause the observed anomaly and also explain other smaller  inconsistencies \cite{Descotes-Genon:2013wba}. 

At present the most stringent constraint on the considered new-physics scenario arises from the mass difference in $B_s$--$\bar B_s$ mixing, which effectively leads to a lower bound of  $\Delta C_9^\ell \sim -1.3$ on the modification in the relevant Wilson coefficient. Negative shifts of this size would induce effects of ${\cal O} (-20\%)$ in $B \to X_s \ell^+ \ell^-$ and ${\cal O} (-25\%)$ in $B^+ \to K^+ \mu^+ \mu^-$ without affecting $B_s \to \mu^+ \mu^-$. Updated  analyses of  the theoretically clean $B \to X_s \ell^+ \ell^-$ mode seem hence more important than ever, and we encourage both the BaBar and the Belle collaborations to analyse their full data sets. In the case of the exclusive channel a better theoretical understanding and a careful assessment of form-factor uncertainties and the impact of contributions from charmonium resonances is needed to fully exploit the power of the $B^+ \to K^+ \mu^+ \mu^-$ constraint. This makes clear that a combined experimental and theoretical effort  is  indispensable to draw more definitive conclusions about the deviations in $B \to K^\ast \mu^+ \mu^-$ as reported by LHCb. 

We also showed that the pattern (\ref{eq:bestfit}) of new-physics effects is correlated to enhancements of ${\cal O} (20 \%)$ in the rates of $B \to K^{(\ast)}, X_s \nu \bar \nu$~(see also  \cite{Altmannshofer:2013foa, Gauld:2013qba}). More than $50 \, {\rm ab}^{-1}$ of SuperKEKB data would however be needed to detect corrections of this size \cite{Aushev:2010bq}.  The considered $Z^\prime$-boson effects have the correct sign to improve the notorious deviations in the $B \to \pi K$ sector, but they are numerically far too small to provide an full explanation.

 Since the large corrections $\Delta C_9^\ell$ require that the $Z^\prime$ boson couples strongly to both the $\bar s b$ and $\bar \mu \mu$ currents, possible physics interpretations of the $B \to K^\ast \mu^+ \mu^-$ anomaly can also be cross-checked with precision measurements of charged-lepton properties. Here the most stringent constraint arises at present from first row CKM unitarity, which gives a bound of $M_{Z^\prime} > 2.7 \, {\rm TeV}$. The upgrade of the JLAB polarized electron beam and the further development of the MAMI facility in Mainz will advance the precision frontier for low-energy parity violation. The Qweak experiment will probe the weak charge of the proton  $Q^p_W$ in electron-proton scattering,  anticipating a relative experimental uncertainty of around $4\%$~\cite{VanOers:2007zz}, while the  P2 experiment aims for a relative precision of $2\%$ \cite{P2}. Using (\ref{eq:dQW}) and (\ref{GVqZN}) these accuracies translate into a reach on $M_{Z^\prime}$ of $2.1 \, {\rm TeV}$ and $2.9  \, {\rm TeV}$, respectively. The goal of the proposed MOLLER experiment is to measure  the weak charge of the electron $Q^e_W$ to about $2\%$  \cite{Moller}. From (\ref{eq:dQWe}) we see that this precision would allow to exclude $M_{Z^\prime}$ values up to $2.9 \, {\rm TeV}$. The next generation of low-energy parity-violation experiments will hence start to become sensitive to the range of $Z^\prime$-boson masses that can already now be probed by first row CKM unitarity. 

Given the  leptophilic character of the $Z^\prime$ boson in the  $3$-$3$-$1$ model with $\beta = -\sqrt{3}$, the most recent resonance searches in the di-lepton invariant mass spectrum at LHC already exclude values $M_{Z^\prime} > 3.9 \, {\rm TeV}$ at 95\% CL.  This bound is expected to significantly improve at the $14 \, {\rm TeV}$ LHC: while it should be possible with $0.3 \, {\rm ab}^{-1}$ of integrated luminosity to discover $Z^\prime$ bosons with masses up to ${\cal O} (5 \, {\rm TeV})$, a HL-LHC delivering $3 \, {\rm ab}^{-1}$ of data would be needed to see evidence for $Z^\prime$ bosons  as heavy as (\ref{eq:MZp}).  An ILC with a CM energy of $500 \, {\rm GeV}$ would be superior in this context and should be able to probe values of $M_{Z^\prime}$ up to  ${\cal O} (10 \, {\rm TeV})$ by a precision measurement of the $e^+  e^- \to \mu^+ \mu^-$ cross section. 

We have seen that achieving a large negative correction $\Delta C_9^\ell$ requires the presence of a $Z^\prime$ boson with a mass in the ballpark of  ${\cal O} (7 \, {\rm TeV})$. Such large values of $M_{Z^\prime}$ are potentially problematic in the $3$-$3$-$1$ scenario with  $\beta = -\sqrt{3}$, since the $Z^\prime$-boson couplings to fermions will develop a Landau pole at high enough energies.  In the original model~\cite{Pisano:1991ee,Frampton:1992wt} with minimal particle content, the relevant couplings become non-perturbative for energy scales of~${\cal O} (5 \, {\rm TeV})$, but the turn-on of non-perturbative dynamics can be delayed by suitably enlarging the particle content of the $3$-$3$-$1$ model \cite{Dias:2004dc, Dias:2004wk}. To which extend such an UV deformation modifies the  physics at low energies is an very interesting question, that is however beyond the scope of the present paper. 

Altogether we believe that the $3$-$3$-$1$ scenario with  $\beta = -\sqrt{3}$ studied in our work is a prototype of an explicit new-physics scenario that can successfully address the $B \to K^\ast \mu^+ \mu^-$ anomaly. The real question is now whether the observed deviations are a manifestation  of beyond the SM physics or a simple result of statistical fluctuations combined with  too optimistic theory errors. The final verdict is still out!

\acknowledgments 

We thank Wolfgang Altmannshofer, Andrzej Buras, Sebastian~J\"ager, Quim Matias and in particular David Straub for useful discussions and communications. A big ``thank you'' goes to Andrzej Buras for pointing out errors in various equations in Section~\ref{sec:PM} of the first version of this manuscript (see also \cite{Buras:2013dea}). The research of R.~G. is supported by an STFC Postgraduate Studentship and F.~G. acknowledges support by the Swiss National Foundation under contract SNF 200021-143781.

\end{document}